\documentstyle[psfig]{mn} 
\title{The intermediate-age open clusters Ruprecht~4,
Ruprecht~7 and Pismis~15}
\author[G. Carraro et al.]        
{G. Carraro$^{1,2}$
\thanks{On leave from Dipartimento di Astronomia, Universit\`a di Padova,
Vicolo Osservatorio 2, I-35122, Padova, Italy.
email: gcarraro@das.uchile cl}, D. Geisler$^{3}$, G. Baume$^{4}$, R. V\'azquez$^{4}$, and A. Moitinho$^{5}$\\
$^1$Departamento de Astronom\'ia, Universidad de Chile, 
Casilla 36-D, Santiago, Chile\\
$^2$Astronomy Department, Yale University, 
P.O. Box 208101, New Haven, CT 06520-8101 , USA\\
$^3$Universidad de Concepci\'on, Departamento de Fisica, 
Casilla 160-C, Concepci\'on, Chile\\
$^4$Facultad de Ciencias Astron\'omicas y Geof\'{\i}sicas de la
UNLP, IALP-CONICET, Paseo del Bosque s/n, La Plata, Argentina\\
$^5$CAAUL, Observat\'orio Astron\'omico de Lisboa, Tapada da Ajuda,
  1349-018 Lisboa, Portugal\\
} 
 
\date{\it Submitted: February 2005} 
\pubyear{2005} 
\begin{document} 
\maketitle 
\title{} 
 
\begin{abstract} 
We report on $BVI$ CCD photometry to $V=22.0$ for 3 fields centered  
on the region of the Galactic
star  clusters Ruprecht~4, Ruprecht~7 and Pismis~15 
and on 3 displaced control fields. 
Ruprecht~4 and Pismis~15 have never been studied before, and we provide
for the first time estimates of their fundamental parameters,
namely, radial extent, age, distance and reddening.
Ruprecht~7 (Berkeley~33) however was studied by Mazur et al. (1993).
We find that the three clusters are all of intermediate age (0.8-1.3
Gyr), and with a metallicity close to or lower than solar.
\end{abstract} 
 

\section{Introduction}
This paper belongs to a series dedicated to the study of the open clusters
population in the third Galactic Quadrant, and aiming at addressing
fundamental questions like the structure of the spiral arms in this quadrant,
and the precise  definition of the Galactic disk radial abundance gradient
outside the solar circle.
A more detailed illustration of the motivations of this project are given 
in Moitinho (2001) and Baume et al (2004).
Here we concentrate on three intermediate-age clusters
(about the age of the Hyades - 600 Myrs - or older), 
Ruprecht~4, Ruprecht~7 and Pismis~15, for which
we provide new photometric data and try to clarify
their nature and to derive the first estimates of their
fundamental parameters.\\
\noindent
The layout of the paper is as follows. Sect.~2 illustrates  
the observation and reduction strategies. 
An analysis of  the geometrical
structure and star counts in the field of the clusters
are presented in Sect.~3, whereas a discussion of
the Color-Magnitude Diagrams (CMD) is performed in Sect.~4.
Sect.~5 deals with the determination of clusters reddening, 
distance and age and,
finally, Sect.~6 summarizes our findings.

\begin{table}
\caption{Basic parameters of the clusters under investigation.
Coordinates are for J2000.0 equinox}
\begin{tabular}{ccccc}
\hline
\hline
\multicolumn{1}{c}{Name} &
\multicolumn{1}{c}{$RA$}  &
\multicolumn{1}{c}{$DEC$}  &
\multicolumn{1}{c}{$l$} &
\multicolumn{1}{c}{$b$} \\
\hline
& {\rm $hh:mm:ss$} & {\rm $^{o}$~:~$^{\prime}$~:~$^{\prime\prime}$} & [deg] & [deg]\\
\hline
Ruprecht~4          & 06:48:59 & -10:31:10 & 222.04 & -5.31\\ 
Ruprecht~7          & 06:57:52 & -13:13:25 & 225.44 & -4.58\\ 
Pismis~15           & 09:34:45 & -48:02:19 & 272.49 & +2.86\\  
\hline\hline
\end{tabular}
\end{table}

\section{Observations and Data Reduction} 
 
$\hspace{0.5cm}$
CCD $BVI$ observations were carried out with the CCD camera on-board
the  1. 0m telescope at Cerro Tololo Interamerican Observatory (CTIO,Chile), on the nights of 
December 13 and 15, 2004. 
With a pixel size of $0^{\prime\prime}.469$,  and a CCD size of 512 $\times$ 512
pixels,  
this samples a $4^\prime.1\times4^\prime.1$ field on the sky.\\
\noindent
The details of the observations are listed in Table~2 where the observed 
fields are 
reported together with the exposure times, the average seeing values and the 
range of air-masses during the observations. 
Figs.~1 to 3 show finding charts in the area of 
Ruprecht~4, Ruprecht~7 and Pismis~15, respectively.

\noindent
The data have been reduced with the 
IRAF\footnote{IRAF is distributed by NOAO, which are operated by AURA under 
cooperative agreement with the NSF.} 
packages CCDRED, DAOPHOT, ALLSTAR and PHOTCAL using the point spread function (PSF)
method (Stetson 1987). 
The two nights turned out to be photometric and very stable, and therefore
we derived calibration equations for all the 130 standard stars
observed during the two nights in the Landolt 
(1992)  fields SA~95-41, PG~0231+051, Rubin~149, Rubin~152,
T~phe and    SA~98-670 (see Table~2 for details).
Together with the clusters, we observed three control fields  20 arcmins
apart from the nominal cluster centers to deal with field star
contamination. Exposure of 600 secs in V and I were secured for these
fields.

\begin{figure} 
\centerline{\psfig{file=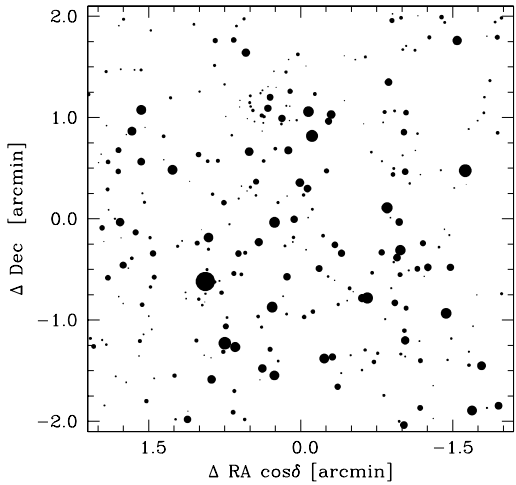,width=\columnwidth}} 
\caption{$V$ Finding charts in the region
of the open cluster Ruprecht~4. 
The size of the dots are proportional to the star magnitude. 
North is up, east on the left, and the covered area is $
4^{\prime}.1 \times 4^{\prime}.1$}
\label{mappa} 
\end{figure} 

\begin{figure} 
\centerline{\psfig{file=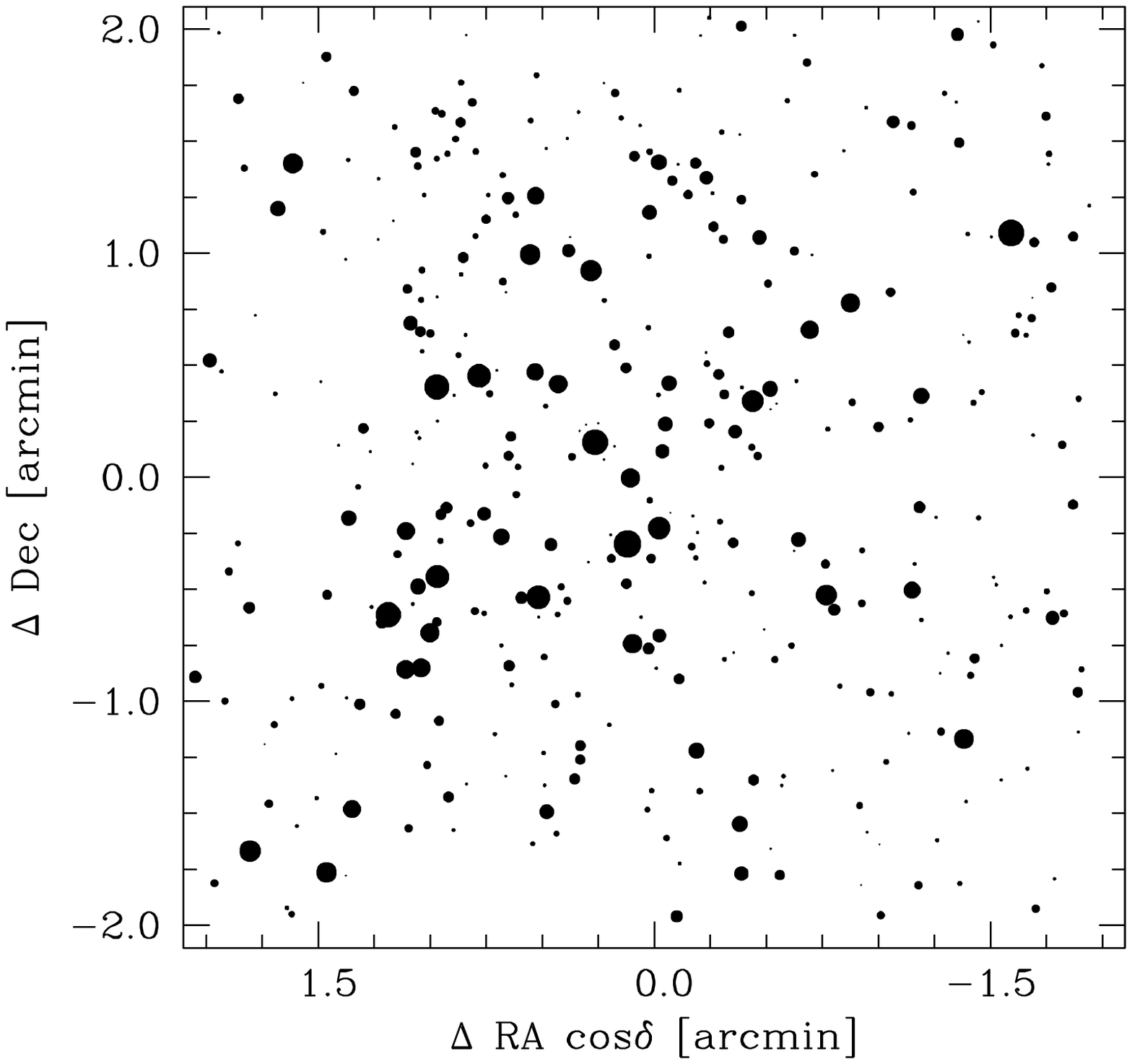,width=\columnwidth}} 
\caption{$V$ Finding chart in the region
of the open cluster Ruprecht~7.
The size of the dots are proportional to the star magnitude. 
North is up, East on the left, and the covered area is $
4^{\prime}.1 \times 4^{\prime}.1$}
\label{mappa} 
\end{figure}

\begin{figure} 
\centerline{\psfig{file=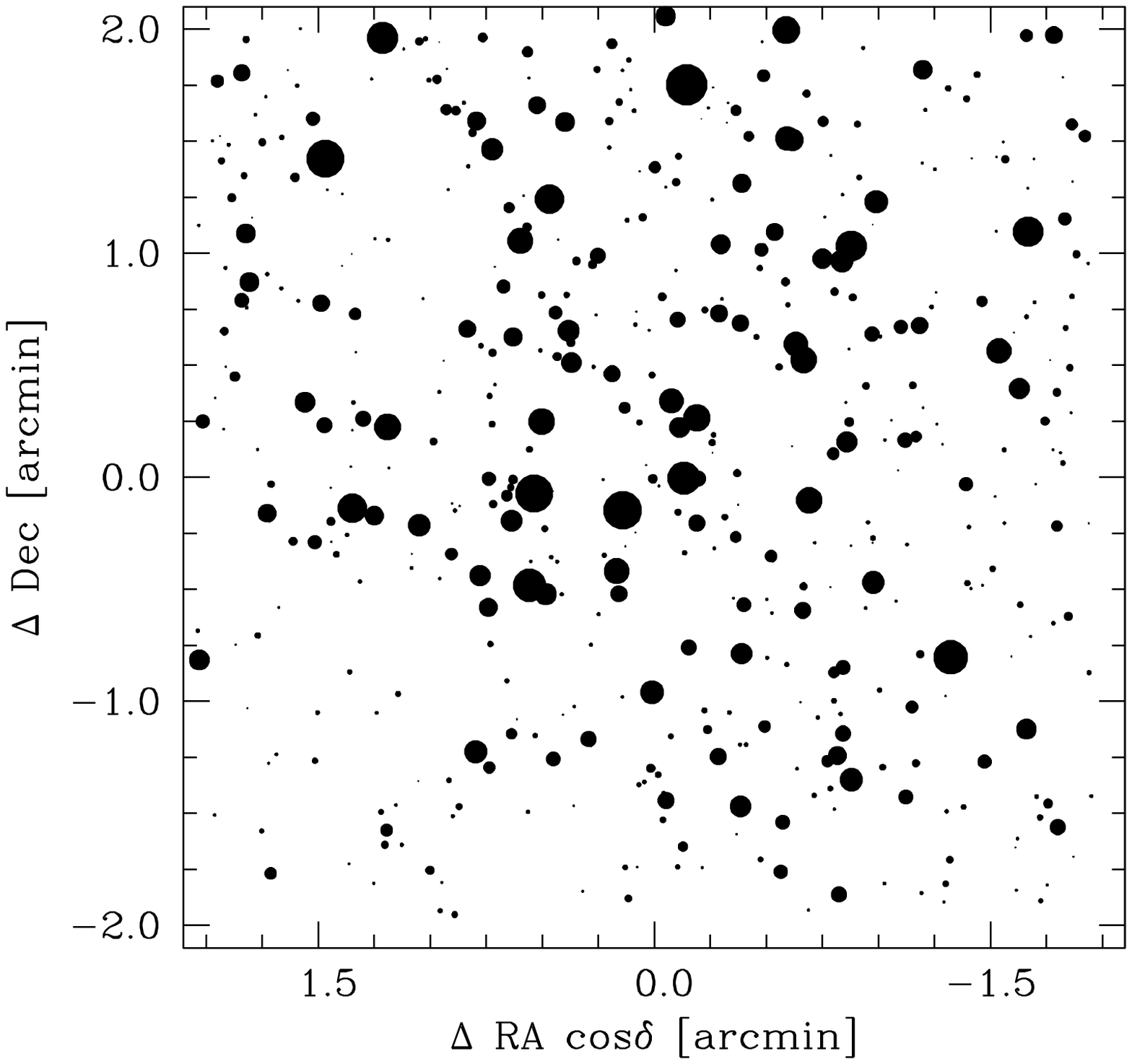,width=\columnwidth}} 
\caption{$V$ Finding chart in the region
of the open cluster Pismis~15. 
The size of the dots are proportional to the star magnitude. 
North is up, East on the left, and the covered area is $
4^{\prime}.1 \times 4^{\prime}.1$}
\label{mappa} 
\end{figure}

\begin{table} 
\fontsize{8} {10pt}\selectfont
\tabcolsep 0.10truecm 
\caption{Journal of observations of Ruprecht~4, Ruprecht~7 and Pismis~15 
and standard star fields (December 13 and 15, 2004).} 
\begin{tabular}{cccccc} 
\hline 
\multicolumn{1}{c}{Field}         & 
\multicolumn{1}{c}{Filter}        & 
\multicolumn{1}{c}{Exposure time} & 
\multicolumn{1}{c}{Seeing}        &
\multicolumn{1}{c}{Airmass}       \\
 & & [sec.] & [$\prime\prime$] & \\ 
\hline 
Ruprecht~4    & B &         120,1200   &   1.2 & 1.12-1.20 \\
              & V &      30,600    &   1.3 & 1.12-1.20 \\ 
              & I &      30,600    &   1.2 & 1.12-1.20 \\
\hline
Ruprecht~7     & B &         120,1200   &   1.2 & 1.12-1.20 \\
              & V &      30,600    &   1.3 & 1.12-1.20 \\ 
              & I &      30,600    &   1.2 & 1.12-1.20 \\
\hline
Pismis~15     & B &         120,1200   &   1.2 & 1.12-1.20 \\
              & V &      30,600    &   1.3 & 1.12-1.20 \\ 
              & I &      30,600    &   1.2 & 1.12-1.20 \\
\hline
SA 98-671     & B &   $3 \times$120   &   1.2 & 1.24-1.26 \\
              & V &   $3 \times$40    &   1.4 & 1.24-1.26 \\ 
              & I &   $3 \times$20    &   1.4 & 1.24-1.26 \\ 
\hline
PG 0231+051   & B &   $3 \times$120   &   1.2 & 1.20-2.04 \\
              & V &   $3 \times$40    &   1.5 & 1.20-2.04 \\ 
              & I &   $3 \times$20    &   1.5 & 1.20-2.04 \\ 
\hline
T Phe         & B &   $3 \times$120   &   1.2 & 1.04-1.34 \\
              & V &   $3 \times$ 40   &   1.3 & 1.04-1.34 \\ 
              & I &   $3 \times$ 20   &   1.3 & 1.04-1.34 \\ 
\hline
Rubin 152     & B &   $3 \times$120   &   1.3 & 1.33-1.80 \\
              & V &   $3 \times$40    &   1.2 & 1.33-1.80 \\ 
              & I &   $3 \times$20    &   1.2 & 1.33-1.80 \\ 
\hline
Rubin 149     & B &   $3 \times$120   &   1.1 & 1.21-1.96 \\
              & V &   $3 \times$40    &   1.2 & 1.21-1.96 \\ 
              & I &   $3 \times$20    &   1.2 & 1.21-1.96 \\ 
\hline
SA 95-41      & B &   $3 \times$120   &   1.2 & 1.05-1.48 \\
              & V &   $3 \times$40    &   1.2 & 1.05-1.48 \\ 
              & I &   $3 \times$20    &   1.1 & 1.05-1.48 \\ 
\hline
\hline
\end{tabular}
\end{table}

\noindent
The calibration equations turned out of be of the form:\\

\noindent
$ b = B + b_1 + b_2 * X + b_3~(B-V)$ \\
$ v = V + v_1 + v_2 * X + v_3~(B-V)$ \\
$ v = V + v_{1,i} + v_{2,i} \times X + v_{3,i} \times (V-I)$ \\
$ i = I + i_1 + i_2 * X + i_3~(V-I)$ ,\\

\noindent

\begin{table} 
\tabcolsep 0.3truecm
\caption {Coefficients of the calibration equations}
\begin{tabular}{ccc}
\hline
$b_1 = 3.465 \pm 0.009$ & $b_2 =  0.25 \pm 0.02$ & $b_3 = -0.145 \pm 0.008$ \\
$v_1 = 3.244 \pm 0.005$ & $v_2 =  0.16 \pm 0.02$ & $v_3 =  0.021 \pm 0.005$ \\
$i_1 = 4.097 \pm 0.005$ & $i_2 =  0.08 \pm 0.02$ & $i_3 =  0.006 \pm 0.005$ \\
$i_1 = 4.097 \pm 0.005$ & $i_2 =  0.08 \pm 0.02$ & $i_3 =  0.006 \pm 0.005$ \\
\hline
\end{tabular}
\end{table}

\noindent
where $BVI$ are standard magnitudes, $bvi$ are the instrumental ones and  $X$ is 
the airmass; all the coefficient values are reported in Table~3.
The standard 
stars in these fields provide a very good color coverage.
The final {\it r.m.s.} of the calibration are 0.039, 0.034 and 0.033 for the B, V and I filters,
respectively.

\noindent
We generally used the third equation to calibrate the $V$ magnitude
in order to get the same magnitude depth both in the cluster
and in the field.\noindent
Photometric errors have been estimated following Patat \& Carraro (2001).
It turns out that stars brighter than  
$V \approx 22$ mag have  
internal (ALLSTAR output) photometric errors lower 
than 0.10~mag in magnitude and lower than 0.18~mag in color, as one can readily see
by inspecting Fig.~3. There the trend of errors in colors and magnitude
are reported against the V mag.\\
\noindent
The final photometric catalog for Ruprecht~4, Ruprecht~7 and Pismis~15 (coordinates,
B, V and I magnitudes and errors)  
consists of 953, 769 and 1007 stars, respectively, and are made 
available in electronic form at the  
WEBDA\footnote{http://obswww.unige.ch/webda/navigation.html} site
maintained by J.-C. Mermilliod.\\

\begin{figure}
\centering
\centerline{\psfig{file=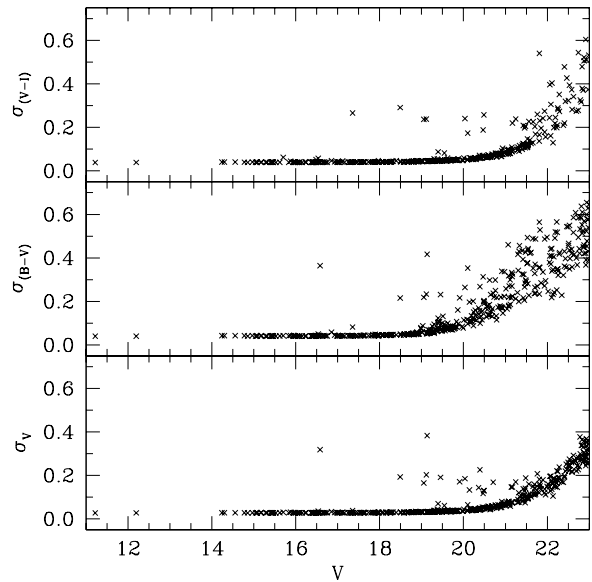,width=\columnwidth}} 
\caption{Trend of  photometric errors in $V$, $(B-V)$  and $(V-I)$
as a function of $V$ magnitude.}
\end{figure}

\begin{figure} 
\centerline{\psfig{file=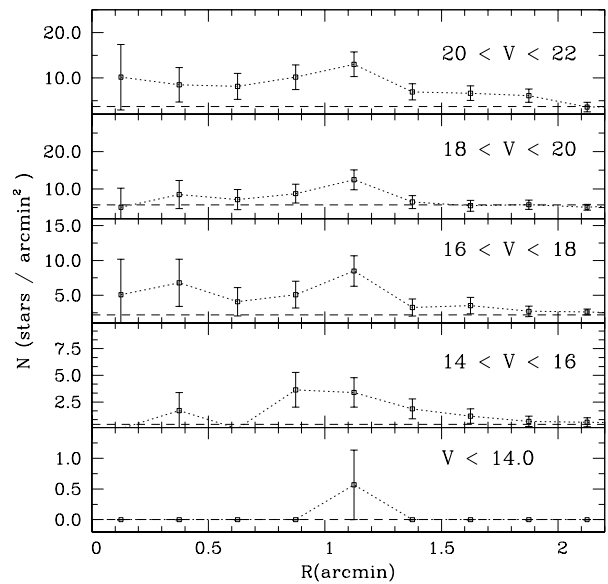,width=\columnwidth}} 
\caption{Star counts in the area of 
Ruprecht~4 as a function of  radius and magnitude. The dashed lines represent
the level of the control field counts estimated in the surroundings
of the cluster in that magnitude range.}
\end{figure}

\section{Star counts and cluster size} 
Since our photometry covers entirely each cluster's area
we performed star counts to obtain
an improved estimate of the clusters size.
We derived the surface stellar density by performing star counts
in concentric rings around the clusters nominal centers (see Table~1)
and then dividing by their
respective area. Poisson errors have also been derived and normalized
to the corresponding area. 
The field star contribution has been derived from the control
field which we secured for each cluster.

\begin{figure} 
\centerline{\psfig{file=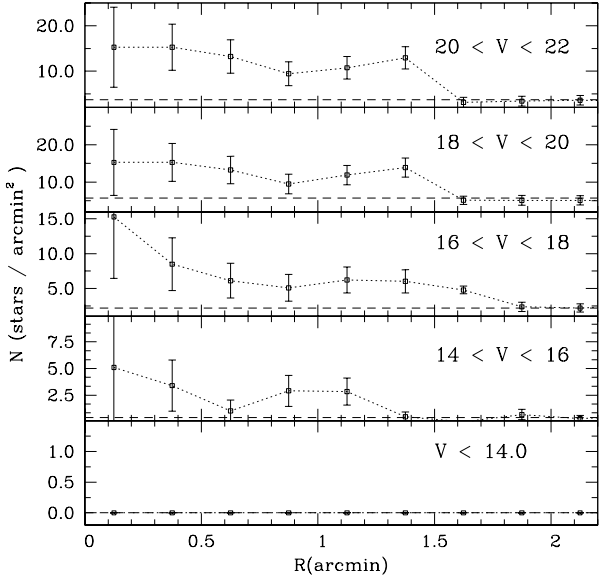,width=\columnwidth}} 
\caption{Star counts in the area of 
Ruprecht~7 as a function of  radius and magnitude. The dashed lines represent
the level of the control field counts estimated in the surroundings
of the cluster in that magnitude range.}
\end{figure}

\begin{figure} 
\centerline{\psfig{file=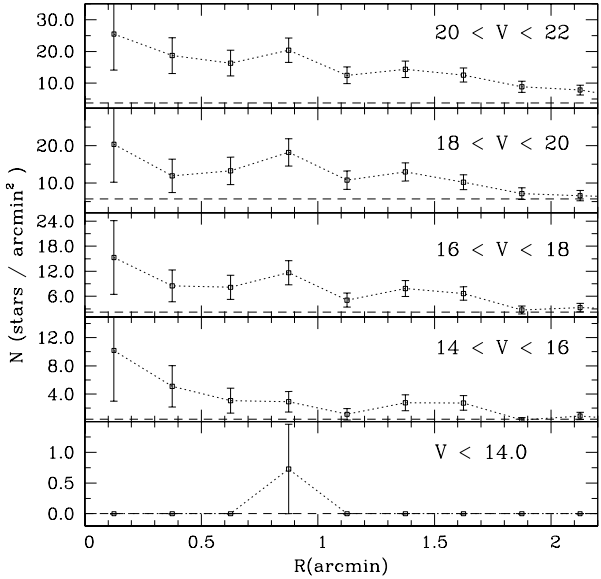,width=\columnwidth}} 
\caption{Star counts in the area of 
Pismis~15 as a function of  radius and magnitude. The dashed lines represent
the level of the control field counts estimated in the surroundings
of the cluster in that magnitude range.}
\end{figure}

{\bf Ruprecht~4}
The final radial density profile for Ruprecht~4 is shown in Fig.~5
as a function of V magnitude.
Clearly,  the cluster does not appear concentrated, and at the nominal
cluster center we can see a deficiency of stars, which on the are hand
populated a sort a ring-like structure (see also Fig.~1) at 1.5 arcmin from the center.
This is not unusual for open clusters, which in many case are sparse objects.\\
The cluster seems to be populated  by stars of magnitude in the range
$15 \leq V \leq 22$, where it clearly emerges from the background.
In this magnitude range the radius is not larger than 1.5 arcmin.
\noindent
We shall adopt the  value of 1.5 arcmin as 
the Ruprecht~4
radius throughout this paper. This estimate is smaller than
the value of 4.5 arcmin reported by Dias et al. (2002) for the cluster
diameter.

{\bf Ruprecht~7}
The final radial density profile for Ruprecht~7 is shown in Fig.~6
as a function of V magnitude.
The cluster seems to be populated  by stars of magnitude in the range
$14 \leq V \leq 22$, where it clearly emerges from the background.
In this magnitude range the radius is not larger than 1.5 arcmin.
\noindent
In conclusion,  we are going to adopt the  value of 1.5 arcmin as 
the Ruprecht~7
radius throughout this paper. This estimate is smaller than
the value of 4.0 arcmin reported by Dias et al. (2002) for the cluster
diameter.

{\bf Pismis~15}
The final radial density profile for Pismis~15 is shown in Fig.~7
as a function of V magnitude.
The cluster seems to be populated  by stars of magnitude in the range
$14 \leq V \leq 22$, where it clearly emerges from the background.
In this magnitude range the radius is not larger than 1.8 arcmin.
\noindent
In conclusion,  we are going to adopt the  value of 1.8 arcmin as 
the Pismis~15
radius throughout this paper. This estimate is in good agreement with
the value of 4.0 arcmin reported by Dias et al. (2002) for the cluster
diameter.\\

\noindent
The estimates we provide for the radius, although reasonable,
must be taken as preliminary.
In fact the size of the CCD is probably too small to derive 
conclusive estimates of the cluster sizes. This is particularly
true in the case Pismis~15, for which the cluster radius we derived
must be considered as a lower limit of the real cluster radius.
In fact, while for Ruprecht~4 and 7 the cluster density profile 
converges toward the field level within the region covered by the CCD,
in the case of Pismis 15 the cluster dominates the star counts - at least
for the faintest stars -
up to the border of the region we covered. Larger field coverage
is necessary in this case to derive a firm estimate of the cluster
radius.

\begin{figure*} 
\centerline{\psfig{file=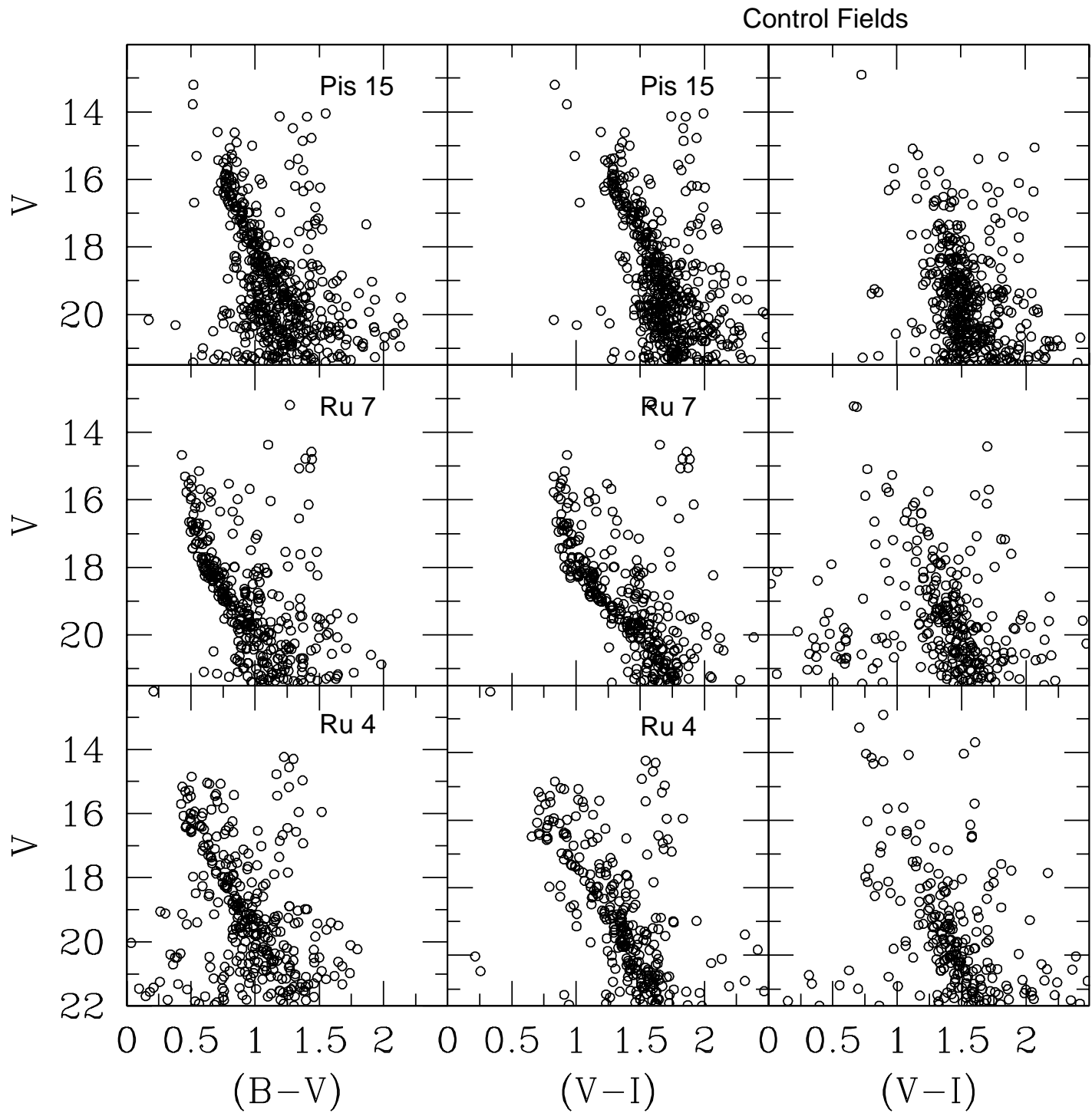}} 
\caption{$V$ vs $(B-V)$ (left panels) and $V$ vs $(V-I)$ (middle panels) CMDs
of Ruprecht~4 (lower panels), Ruprecht~15 (middle panels) and
Pismis~15 (upper panels) and corresponding control fields (right panels).
We include all stars in each field.}
\end{figure*}


\section{The Colour-Magnitude Diagrams} 
In Fig.~8 we present the CMDs we obtained for the three clusters
under investigation.
In this figure the open cluster Ruprecht~4 is shown  together
with the corresponding control field in the lower panels,
whereas Ruprecht~7 and Pismis~15 are presented in the
middle and upper panels, respectively. The control fields
help us to better interpret these CMDs, which are clearly
dominated by foreground star contamination.

\noindent
{\bf Ruprecht~4}.
This cluster is presented in the lower panels of Fig.~8.
It exhibits a Main Sequence (MS) extending from
V=16, where the Turn Off Point (TO) is located down
to V=22. This MS is significantly wide, a fact that we
ascribe to the increasing photometric error at increasing magnitude,
the field star contamination, 
and to the presence of a sizeable binary star population,
which mainly enlarge the MS toward red colors.
In particular the effect of a significant binary population
is known to affect the MS and TO shape, which appear at first glance
confused (see for comparison the CMDs in the middle
and upper panels; for a reference see also Meynet et al. 1993).

However, the reality of this cluster seems to be secured by the shape of
the MS with respect to the control field MS, which 
population sharply decrease at V = 18.
Besides, the cluster MS is significantly bluer and more tilted
than the field MS, which derives from the superposition
of stars of different reddening located at all distances between
the cluster and the Sun.
Another interesting evidence it the possible presence of  a clump
of stars at V=15, which does not have a clear counterpart
in the field, and which makes the cluster an intermediate-age one.
In fact if we use the age calibration from Carraro \& Chiosi (2004),
for a $\Delta V$ (say the magnitude difference between
the red clump ant the TO) of 1 mag, we infer an age around 1 billion
year. This estimate does not take into account the cluster metallicity,
and therefore is simply a guess. In the following we shall
provide a more robust estimate of the age through a detailed
comparison with theoretical isochrones.

{\bf Ruprecht~7}.
The open cluster Ruprecht~7 is presented in the middle panels
of Fig.~8.
The interpretation of this CMD seems much easier than the previous
one.
Here the MS is more evident, the TO is located at $V \approx$ 15.5, and the clump
at $V \approx$15, thus implying a rough estimate for the age around 0.5 billion
year. The overall morphology of the CMDs is so different from the
field CMD that it leaves no doubt on the cluster reality.
Also in this case the distribution of the stars in the red edge of the MS
suggests a probably binary star population.
Both Ruprecht~4 and Ruprecht~7 has a clump mean magnitude
V=15. According to Salaris \& Girardi (2002), this implies a common
distance modulus (m-M) of about 15 mag.

\begin{figure*} 
\centerline{\psfig{file=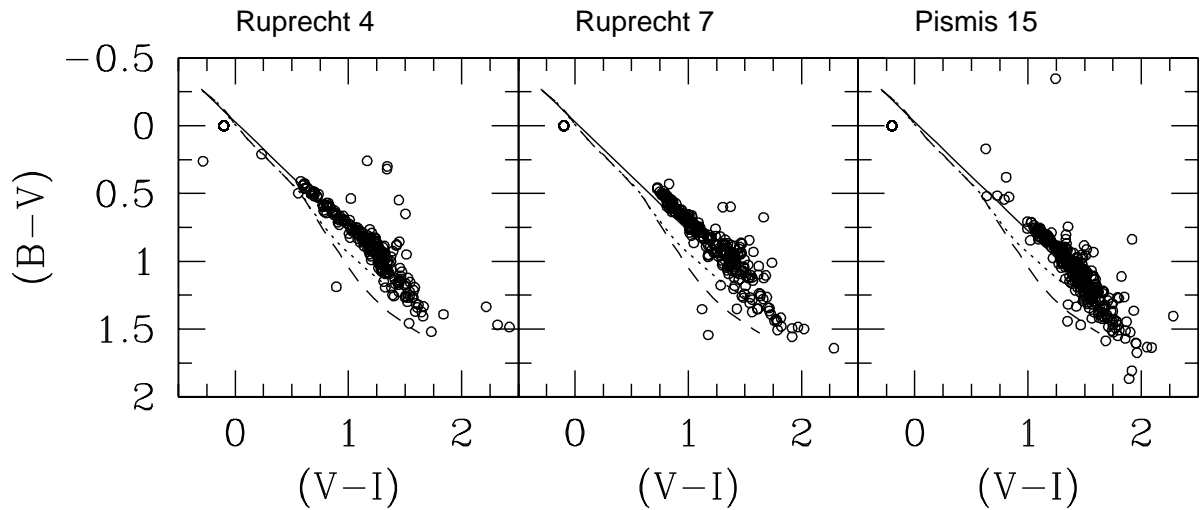,width=16cm}} 
\caption{$B-V$ vs $V-I$ diagram for the three clusters. The solid line
is the normal reddening law, whereas dashed and dotted lines are the
luminosity class III and V ZAMS, respectively. See text for details.}
\end{figure*} 

{\bf Pismis~15}.
The open cluster Pismis~15 is finally presented in the upper panels
of Fig.~8.
The TO is located at $V \approx$ 16, and the clump
at $V \approx$ 14.5, thus implying a rough estimate for the age around 1.5 billion
year. The overall morphology of the CMDs is also in this case
very different from the
field CMD so this is a bona-fide cluster.
A word of caution is mandatory because the precise clump
position is affected by field star contamination in the form of a tail
of stars which connects the clump with the MS at V = 18.
Again, the control field helps us to interpret the CMD.
Indeed also in the control field there is a vertical sequence
detaching from the MS at V=18, which however stops at V = 15,
supporting the interpretation of the 5-6 bright red stars as  clump
stars with mean V =14.5. This suggests that Pismis~15
might be somewhat closer to the Sun than the two other clusters.

\section{Deriving clusters' fundamental parameters}
In this section we are going to perform a detailed comparison
of the star distribution in the clusters CMDs with theoretical
isochrones. We adopt in this study the Padova library from
Girardi et al. (2000).
This comparison is clearly not an easy exercise. In fact
the detailed shape and position of the various features
in the CMD (MS, TO and clump basically) depends mostly
on age and metallicity, and then also on reddening and distance.
The complex interplay between the various parameters is well
known, and we refer to 
Chiosi et al. (1992) and Carraro (2005)
as nice examples of the underlying technique.\\
Our basic strategy is to survey different age and metallicity
isochrones attempting to provide the best fit of all the CMD
features both in the $V$ vs $(B-V)$ and in the $V$ vs $(V-I)$ CMD.\\
Besides, to further facilitate the fitting procedure
we shall consider only the stars which lie  within
the cluster radius as derived in Sect.~3.
\noindent
Therefore, in the series of Figs.~9 to 11 we shall present the best fit
we were able to achieve.
Together with the best fit, we could make estimates of 
uncertainties in the basic parameters derivation. These uncertainties simply
reflect the range in the basic parameter 
which allow a reasonable fit to the clusters CMDs.
Error estimates are reported in Table~4.\\

\noindent
Finally, to derive clusters' distances from reddening and apparent distance modulus,
a reddening law must be specified. In Fig.~9 we show that the normal extinction law
is valid for all the clusters, and therefore we shall us the relation
$Av = 3.1 \times E(B-V)$ to derive clusters' distances. In details, the solid line
in Fig.~9 is the normal extinction $E(V-)=1.245 \times E(B-V)$ law 
from Cousin (1978), whereas the dashed and dotted lines are the luminosity class $III$
and $V$ ZAMS, respectively.

{\bf Ruprecht~4}.
The isochrone solution for this cluster is discussed in Fig.~10.
We obtained the best fit for an age of 800 million years and a metallicity
Z=0.008, quite low for an open cluster of this age. The inferred
reddening and apparent distance modulus are E(B-V)=0.36 (E(V-I)=0.50)
and (m-M)=14.6, respectively. As a consequence the cluster
possesses a heliocentric distance of 4.9 kpc, and
is located at a Galactocentric distance of 12.0 kpc, assuming
8.5 kpc as the distance of the Sun to the Galactic Center.
Interestingly, this cluster appears to be relatively young but very
metal poor. This is not an isolated case, and Moitinho et al. (2005)
reported on a similar case, the star cluster NGC~2635.
The overall fit is very good, the detailed shape of the MS
and TO are nicely reproduced, and the color of the clump as well.

{\bf Ruprecht~7}
The isochrone solution for this cluster is discussed in Fig.~11.
We obtained the best fit for an age of 800 million years and a metallicity
Z=0.019. This metallicity is larger than the one proposed by Mazur et al. (1993),
which was based on a comparison with isochrone from Bertelli et al. (1994).
The inferred
reddening and apparent distance modulus are E(B-V)=0.30 (E(V-I)=0.47)
and (m-M)=15.0, respectively. As a consequence the cluster
lies at 6.5 kpc from the Sun, and 
is located at a Galactocentric distance of 13.8 kpc
toward the anti center direction. 
The overall fit is very good also in this case, the detailed shape of the MS
and TO are nicely reproduced, and the color of the clump as well.\\
With respect to Mazur et al. (1993) study, we obtain a larger distance.
However, both the age and the reddening are compatible with that study.
We believe that the larger distance is due to the different isochrones set
and mostly to the higher metallicity here adopted for this cluster.

{\bf Pismis~15}.
The isochrone solution for this cluster is discussed in Fig.12.
We obtained the best fit for an age of 1300 million year and a metallicity
Z=0.008. The inferred
reddening and apparent distance modulus are E(B-V)=0.53 (E(V-I)=0.88)
and (m-M)=14.0, respectively. Therefore the cluster has a heliocentric distance
of 2.9 kpc,
and
is located at a Galactocentric distance of 8.8 kpc. 
The overall fit is very good also in this case, the detailed shape of the MS
and TO are nicely reproduced, and the color of the clump as well.

\begin{table*}
\caption{Fundamental parameters of the studied clusters. The coordinates system is such that
the Y axis connects the Sun to the Galactic Center, while the X axis is perpendicular to that.
Y is positive toward the Galactic anticenter, and X is positive in the first and second Galactic quadrants (Lynga 1982)}
\fontsize{8} {10pt}\selectfont
\begin{tabular}{cccccccccccc}
\hline
\multicolumn{1}{c} {$Name$} &
\multicolumn{1}{c} {$Radius(arcmin)$} &
\multicolumn{1}{c} {$E(B-V)$}  &
\multicolumn{1}{c} {$E(V-I)$}  &
\multicolumn{1}{c} {$(m-M)$} &
\multicolumn{1}{c} {$d_{\odot}$} &
\multicolumn{1}{c} {$Y(kpc)$} &
\multicolumn{1}{c} {$X(kpc)$} &
\multicolumn{1}{c} {$Z(kpc)$} &
\multicolumn{1}{c} {$R_{GC}(kpc)$} &
\multicolumn{1}{c} {$Age(Gyr)$} & 
\multicolumn{1}{c} {Metallicity}\\
\hline
Ruprecht~4 &  1.5 & 0.36$\pm$0.1 & 0.50$\pm$0.1  & 14.6$\pm$0.2 & 4.9 & -3.6 & -3.3 & -0.45 & 12.0 &0.8$\pm$0.2 & 0.008$\pm$0.005\\
Ruprecht~7 &  1.5 & 0.30$\pm$0.1 & 0.47$\pm$0.1  & 15.0$\pm$0.2 & 7.0 & -4.6 & -4.6 & -0.50 & 13.8 &0.8$\pm$0.3 & 0.019$\pm$0.002\\
Pismis~15  &  1.8 & 0.53$\pm$0.1 & 0.88$\pm$0.1  & 14.0$\pm$0.2 & 2.9 &  0.1 & -2.9 &  0.10 &  8.8 &1.3$\pm$0.3 & 0.008$\pm$0.002\\
\hline
\end{tabular}
\end{table*}

\begin{figure} 
\centerline{\psfig{file=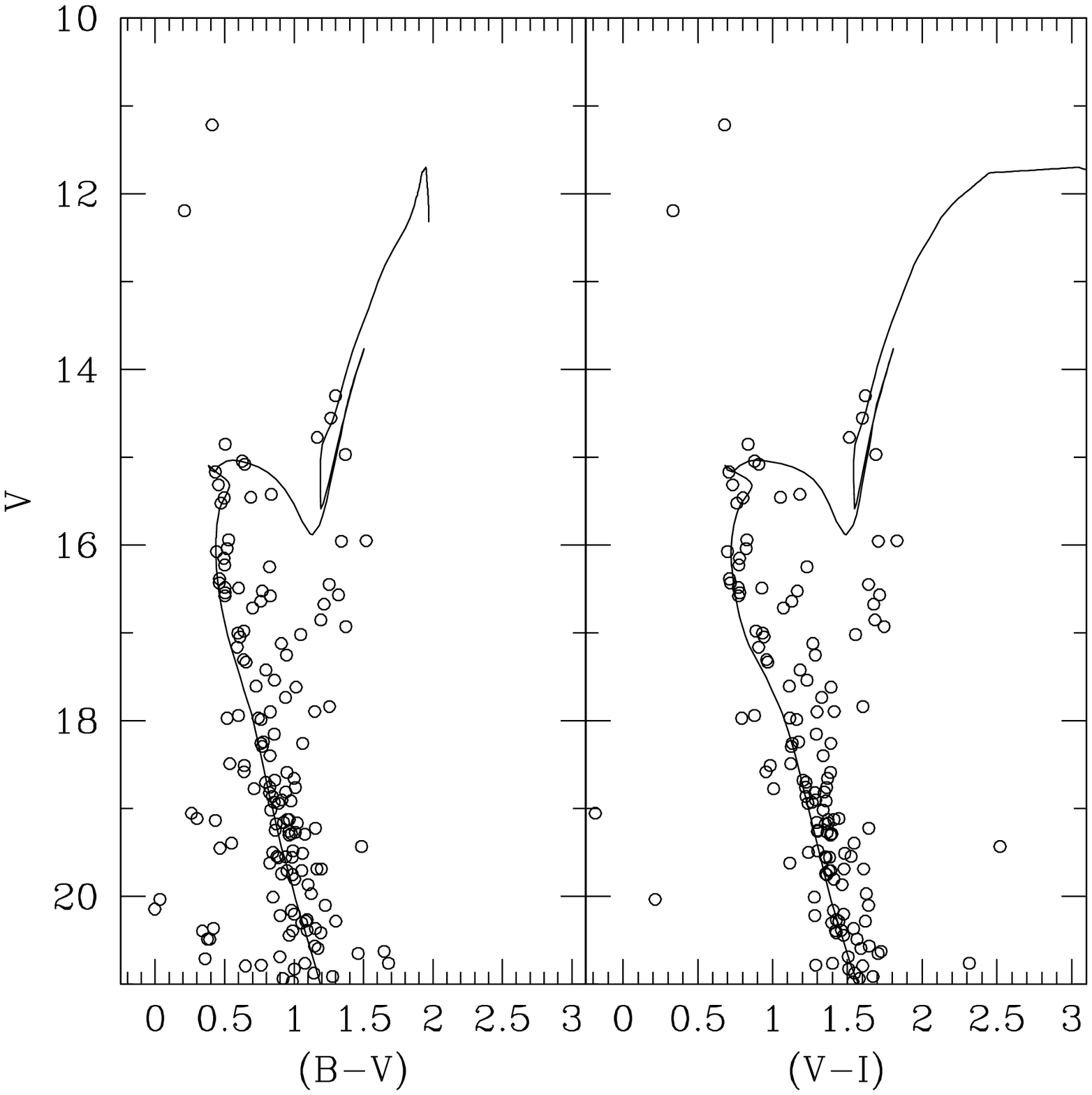,width=\columnwidth}} 
\caption{Isochrone solution for Ruprecht~4. The isochrone
is for the age of 800 million year and metallicity Z=0.008.
The apparent distance modulus is (m-M)=14.6, and the reddening
E(B-V)=0.36 and E(V-I)=0.50. See text for more details. Only stars within the derived radius are shown. }
\end{figure}

\begin{figure} 
\centerline{\psfig{file=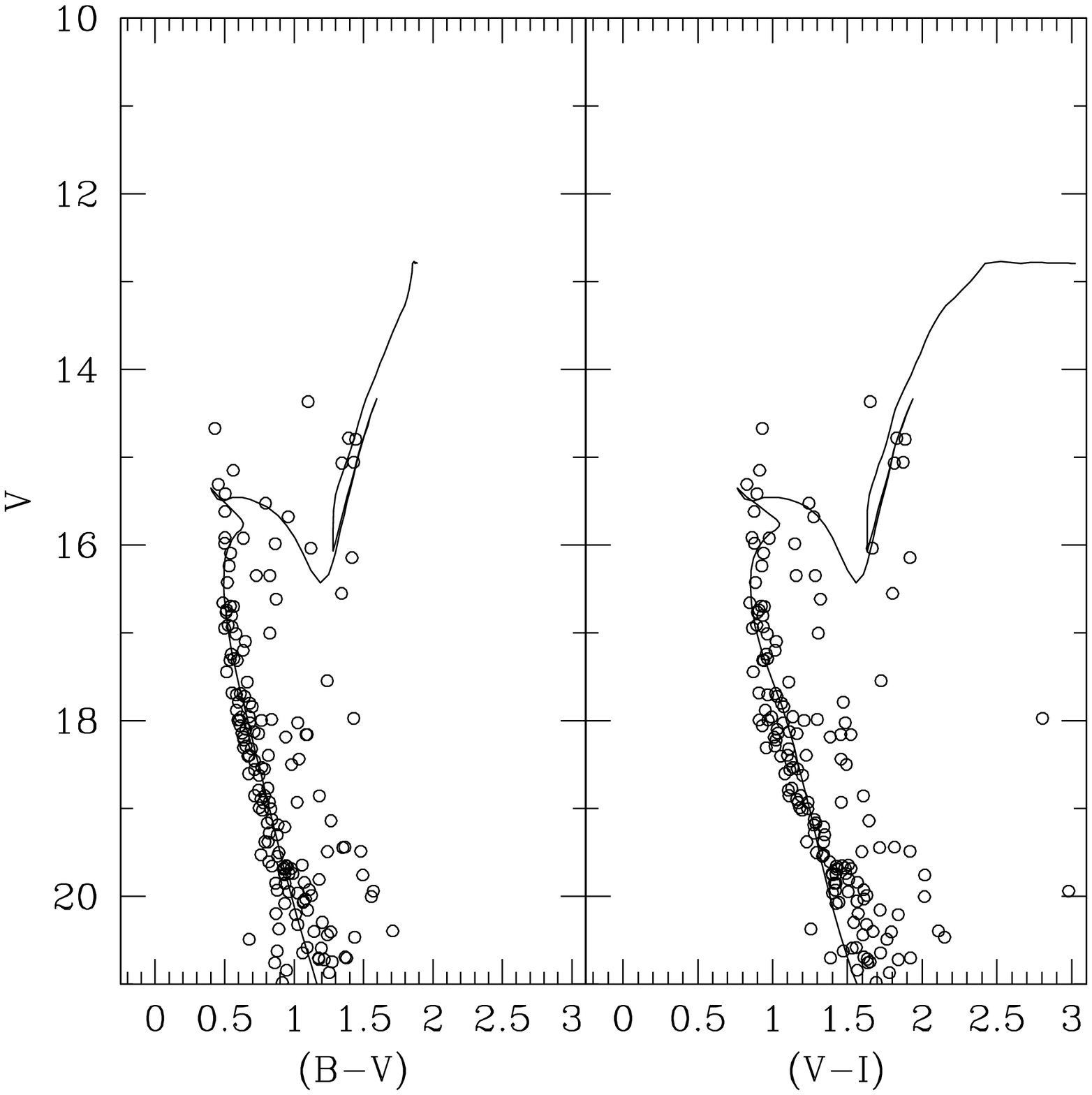,width=\columnwidth}} 
\caption{Isochrone solution for Ruprecht~7. The isochrone
is for the age of 800 million year and metallicity Z=0.019.
The apparent distance modulus is (m-M)=15.0, and the reddening
E(B-V)=0.30 and E(V-I)=0.47. See text for more details. Only stars within the derived radius are shown. }
\end{figure} 

\begin{figure} 
\centerline{\psfig{file=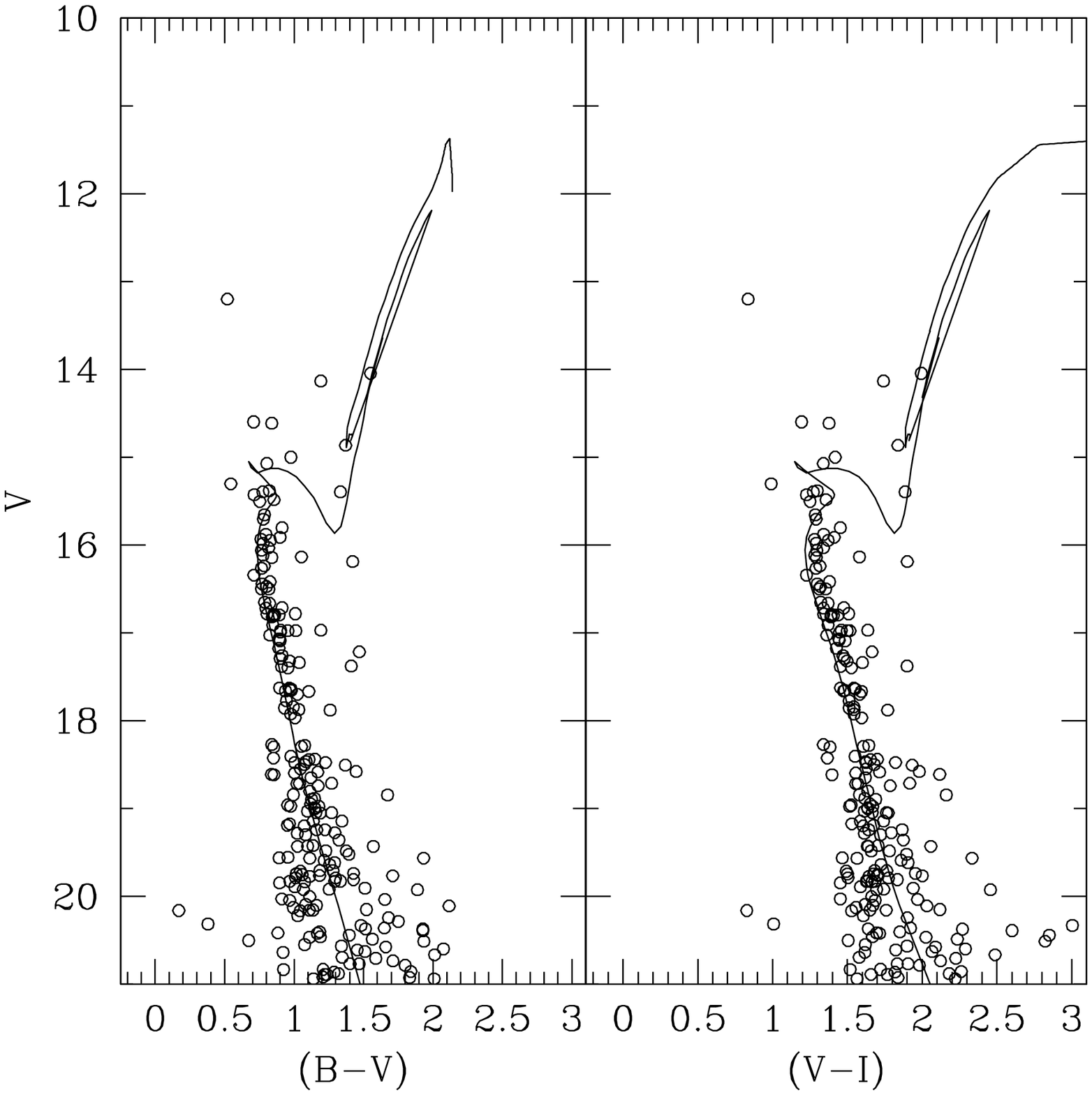,width=\columnwidth}} 
\caption{Isochrone solution for Pismis~15. The isochrone
is for the age of 1300 million year and metallicity Z=0.008.
The apparent distance modulus is (m-M)=14.0, and the reddening
E(B-V)=0.53 and E(V-I)=0.88. See text for more details. 
Only stars within the derived radius are shown.}
\end{figure}

\section{Conclusions}
We have presented the first CCD $BVI$ photometric study of the 
star clusters Ruprecht~4, Ruprecht~7 and Pismis~15. 
The CMDs we derive allow us to 
infer estimates of the clusters' basic parameters, which
are summarized in Table~4.\\
\noindent
In detail, the fundamental findings of this paper are:
 
\begin{description} 
\item $\bullet$ the best fit reddening estimates support
within the errors a normal extinction law toward the three clusters;
\item $\bullet$ Ruprecht~4 is an intermediate age open cluster
whose position is compatible with the cluster belonging to the local spiral arm;
\item $\bullet$ Ruprecht~7 is a Hyades age cluster, although
much poorer in metal than that cluster, and it is located at almost
14.0 kpc in the anti-center direction, thus being one of the most
distant open cluster in the sense of the Galactocentric distance;
\item $\bullet$ Pismis~15 is an old open cluster located inside
the solar ring, with a half than solar metal abundance.
The distance we derived and its Galactic coordinates imply
that the cluster probably belongs to the Carina spiral arm.
\end{description}

\noindent
All these clusters are very interesting in the context of the
chemical evolution of the Galactic disk
(Geisler et al. 1992, Carraro et al. 1998, Friel et al 2002).\\
They may be very useful
to trace the slope of the Galactic disk radial abundance gradient,
which is very poorly sampled in the age range between 0.5 and 2
Gyr (see Friel et al 2002).
If we make use of the provisional photometric estimates
presented in this paper, we obtain that 
all the clusters
basically confirm the most recent slope of the gradient as derive by 
Friel et al. (2002). \\

\noindent
Further studies therefore should concentrate on the confirmation of the clusters'
metal content by means of a detailed abundance analysis of the 
clump stars.

\noindent

\section*{Acknowledgements} 
The observations presented in this paper have been carried out at 
Cerro Tololo Interamerican Observatory CTIO (Chile).
CTIO is operated by the Association of Universities for Research in Astronomy,
Inc. (AURA), under a cooperative agreement with the National Science Foundation
as part of the National Optical Astronomy Observatory (NOAO).
The work of G.C. is supported by {\it Fundaci\'on Andes}.
D.G. gratefully acknowledges support from the Chilean
{\sl Centro de Astrof\'\i sica} FONDAP No. 15010003.
This work has been also developed in the framework of 
the {\it Programa Cient\'ifico-Tecnol\'ogico Argentino-Italiano SECYT-MAE
C\'odigo: IT/PA03 - UIII/077 - per\'iodo 2004-2005}.
This study made use of Simbad and WEBDA databases.


\begin{thebibliography}{} 
\bibitem{} Baume G., Moitinho A., Giorgi E., Carraro G., Vazquez R.,
           2004, A\&A 417, 961
\bibitem{} Bertelli G., Bressan A., Chiosi C., Fagotto F., Nasi E., 1994
           A\&AS 106, 275
\bibitem{} Carraro G., Chiosi C., 1994, A\&A 287, 761
\bibitem{} Carraro G., Ng Y.K, Portinari L., 1998, MNRAS 296, 1045
\bibitem{} Carraro G., 2005, ApJ 621, L61
\bibitem{} Chiosi C., Bertelli G., Bressan A. 1992, ARA\&A 30, 235
\bibitem{} Cousin A.W.J., 1978, MNSSA 37, 42
\bibitem{} Dias W.S., Alessi B.S., Moitinho A., Lepine J.R.D., 2002, 
           A\&A 389, 871 
\bibitem{} Friel E.D., Janes K.A., Tavarez M., Scott J., Katsanis R., Lotz J.,Hong L. \& Miller  N. 2002, AJ 124, 2693
\bibitem{} Geisler D., Claria J.J., Minniti D., 1992, AJ 104, 1892
\bibitem{} Girardi L., Bressan A., Bertelli G., Chiosi C., 2000, 
           A\&AS 141, 371 
\bibitem{} Landolt A.U., 1992, AJ 104, 340
\bibitem{} Lynga G., 1982, A\&A 109, 213
\bibitem{} Mazur B., Kaluzny J., Krzeminski W., 1993, MNRAS 265, 405
\bibitem{} Meynet G., Mermilliod J-C.. \& Maeder A., 1993, A\&AS 98, 477
\bibitem{} Moitinho A., 2001, A\&A  370, 436
\bibitem{} Moitinho A., Carraro G., Baume G., Vazquez R., 2005,
           A\&A submitted
\bibitem{} Patat F., Carraro G., 2001, MNRAS 325, 1591 
\bibitem{} Salaris M., Girardi L., 2002, MNRAS 337, 332
\bibitem{} Schmidt-Kaler Th. 1982, Landolt-B\"ornstein, Numerical data and 
           Functional Relationships in Science and Technology, New Series, Group VI,
           Vol. 2(b), K. Schaifers and H.H. Voigt Eds., Springer Verlag, Berlin, p.14
\bibitem{} Stetson P.B. 1987, PASP 99, 191
\end{thebibliography}
\end{document}